\documentclass[aps,prb,reprint,groupedaddress]{revtex4-1}
\usepackage{graphicx}
\usepackage[utf8]{inputenc}
\usepackage[english]{babel}
\usepackage{color}
\usepackage{soul}
\usepackage[section]{placeins}

\graphicspath{{Images/}}
\usepackage{dcolumn}
\usepackage{xcolor}
\usepackage{bm}
\usepackage{hyperref}
\usepackage{cleveref}%

\begin{document}




\title{Pinning of Domain Walls in thin Ferromagnetic Films}


\author{V. Jeudy}
\email[]{vincent.jeudy@u-psud.fr}
\affiliation{Laboratoire de Physique des Solides, Universit\'e Paris-Sud, Universit\'e Paris-Saclay, CNRS, UMR8502, 91405 Orsay, France.}
\author{R. D\'{\i}az Pardo}
\affiliation{Laboratoire de Physique des Solides, Universit\'e Paris-Sud, Universit\'e Paris-Saclay, CNRS, UMR8502, 91405 Orsay, France.}

\author{W. Savero Torres}
\affiliation{Laboratoire de Physique des Solides, Universit\'e Paris-Sud, Universit\'e Paris-Saclay, CNRS, UMR8502, 91405 Orsay, France.}
\author{S. Bustingorry}
\affiliation{CONICET, Centro At\'omico Bariloche, 8400 San Carlos de Bariloche, R\'{\i}o Negro, Argentina.}
\author{A. B. Kolton}
\affiliation{CONICET and Instituto Balseiro (UNCu), Centro At\'omico Bariloche, 8400 S.C. de Bariloche, Argentina.}



\date{\today}

\begin{abstract}

We present a quantitative investigation of magnetic domain wall pinning in thin magnets with perpendicular anisotropy. 
A self-consistent description exploiting the universal features of the
depinning and thermally activated sub-threshold creep regimes 
observed in the field driven domain wall velocity, 
is used to determine the effective pinning parameters controlling the 
domain wall dynamics: the effective height of pinning barriers, the depinning threshold, and the velocity at depinning. 
Within this framework, the analysis of results published in the literature 
allows for a quantitative comparison of
pinning properties for a set of magnetic materials in a wide temperature range.
On the basis of scaling arguments, the microscopic parameters controlling the pinning: the correlation length of pinning, the collectively pinned domain wall length (Larkin length) and the strength of pinning disorder, are estimated from the effective pinning and the micromagnetic parameters. 
The analysis of thermal effects reveals a crossover between different pinning length scales and strengths at low reduced temperature.

\end{abstract}

%
\pacs{75.78.Fg Dynamics of magnetic domain structures, 68.35.Rh, 64.60.Ht, 05.70.Ln, 47.54.-r,75.50.Pp: Magnetic semiconductors}

\maketitle

\section{Introduction}

A major source of hysteresis in ferromagnets~\cite{hubert_schafer} is the pinning of magnetic domain walls (DWs), which impedes their free motion when driven by an applied magnetic field or a spin current. For a strong pinning, the DWs follow the shape of material defects and magnetization reversal results from percolation processes of magnetic domains~\cite{attane_prl_2004_FePt_percolation}. Weak random pinning also results in important effects: the competition with DW elasticity and thermal activation produces stochasticity~\cite{Kim_PRL2003_avalanches}, domain wall roughness~\cite{lemerle_PRL_1998_domainwall_creep,Moon_PRL_2013}, and dramatically modifies the driven dynamics at small field and current~\cite{lemerle_PRL_1998_domainwall_creep,curiale_prl_2012_spin_drift,duttagupta_natphys_2015}. 
 Weak pinning may result from spatial fluctuations of domain wall energy associated to inhomogeneous thickness in ultrathin metallic films~\cite{lemerle_PRL_1998_domainwall_creep}, or concentration of magnetic atoms in ferromagnetic semiconductors~\cite{lemaitre2008}. As pinning impedes to reach the high velocity flow regimes, several attempts have been proposed to reduce the pinning strength finding low pinning materials~\cite{herrera-diez_apl_2015} and to engineer the pinning properties using light-ion irradiation~\cite{ferre_physStatSolidi_2004,cayssol_apl_2005,franken_jpcm_2012,li_IEEEtrans_2012,herrera-diez_apl_2015} in ultrathin films, or coupling with another magnetic layer~\cite{metaxas_PRL_10_dynamics_coupled_layers}.
 Interestingly, the engineering of pinning is also important for superconducting materials~\cite{kwok_2016,koshelev_2011} and a large variety of methods were developed to enhance the pinning strength on vortices. 
Understanding of the pinning of elastic objects, among which domain walls in thin ferromagnets is a paradigmatic example, is thus of broad interest.

How weak pinning and thermal fluctuations affect the glassy dynamics of domain walls is a critical issue for potential applications based on the controlled motion of domain walls~\cite{parkin_science_2008_race_track,franken_natmat_2012} and for understanding the physics of phenomena as the interaction of spin current with DW or the contribution of the Dzyaloshinskii-Moriya interaction to DW dynamics~\cite{je_prb_2013_DMI}. 
However, going beyond qualitative comparisons between pinning properties of different materials remains challenging. A quantitative framework would be particularly welcomed for a better understanding of DW pinning in thin ferromagnetic films.

%

The pinning dependent DW dynamics combines both universal and material dependent behaviors, which are not straightforward to disentangle. A depinning magnetic field threshold $H_d$ separates the pinning dependent thermally activated so-called creep regime ($H<H_d$) from the depinning transition ($H \geq H_d$) and the flow regime ($H \gg H_d$). Until now, almost all the analysis of experiments on DW dynamics in the creep regime are based on the seminal work of Lemerle {\it et al.}~\cite{lemerle_PRL_1998_domainwall_creep}. In this paper, it was shown that the magnetic field driven DW dynamics can be modeled by the motion of an elastic line in a weakly disordered medium~\cite{chauve2000}.
More precisely, the measured and the predicted creep exponent $\mu$, deduced from the velocity law $v \sim \exp (H^{-\mu})$ and the roughness exponent $\zeta=2/3$ as $\mu=(2 \zeta -1)/(2-\zeta)=1/4$, respectively, were found in good agreement thus attesting the universal nature of DW creep motion. However, those predictions are only valid in the limit $H \rightarrow 0$, which restricts the analysis of domain wall motion to the low drive creep regime. 
%

Recently, the glassy domain wall dynamics was investigated beyond the zero drive limit, and it was shown that the universal creep regime extends up to the depinning threshold~\cite{jeudy_PRL_2016_energy_barrier}. The depinning transition was also found to present universal behaviors~\cite{diaz_PRB_2017_depinning}. Their analysis was pushed beyond the usual asymptotic power law variations. The universal functions describing the creep and depinning regimes could be extracted from experimental results obtained for different materials and temperatures. Moreover, it was shown that both regimes can be described self-consistently using only three parameters absorbing all the intrinsic temperature and material dependent pinning properties. These effective pinning parameters are an effective pinning barrier height $k_BT_d$, where $k_B$ is the Boltzmann constant, a depinning threshold $H_d$ and a depinning velocity $v_T$~\cite{diaz_PRB_2017_depinning}. 
As the latter are directly related to the physics at the so-called Larkin regime of an elastic string in a random medium~\cite{Larkin_1970,larkin1979pinning}, they can be used to bridge between the non-trivial macroscopic universal behavior such as the collective creep and depinning phenomena, to the micromagnetic level of description at which domain walls and their pinning to inhomogeneities emerge.
This situation is analogous to the case of vortex pinning in superconductors, where the Larkin regime bridges between the  macroscopic scale 
and the Ginzburg-Landau continuum description for which vortices are described by well defined pinned elastic objects~\cite{kwok_2016,blatter_vortex_review}. 


The aim of this paper is to understand better the correlations between the material and temperature dependent pinning parameters controlling the glassy DW dynamics and the microscopic origins of DW pinning.  
To this end we exploit the self-consistent ``top-down'' approach~\cite{diaz_PRB_2017_depinning} described above, starting from the identification of universal features in the driven DW glassy dynamical regimes. We deduce a "map" of the material and temperature dependent effective pinning parameters ($T_d$, $H_d$, and $v_T$) controlling DW velocity. We develop a model providing scaling relations between the effective pinning parameters, the micromagnetic parameters (the saturation magnetization $M_s$ and the domain wall surface energy $\sigma$, the domain wall thickness parameter $\Delta$ and the Gilbert damping factor $\alpha$), and the microscopic pinning parameters characterizing the weak pinning disorder (which are the pinning strength $f_{pin}$ and the correlation length of the disorder $ \xi $). The model is used to estimate the microscopic pinning parameters, which are not directly accessible experimentally. A strong modification of domain wall pinning properties at low temperature is evidenced. This work opens a way to a systematic quantitative analysis of magnetic domain wall pinning engineering, which remains however beyond the scope of this paper.
 

%
The organization of the paper is the following. 
The Section~\ref{sec:DWdynamics} discusses DW dynamics: it starts from a qualitative description and extends to the self-consistent modeling, which is used for the extraction of pinning parameters controlling creep and depinning regimes of the velocity.
The Section~\ref{sec:DWpinning} presents a set of pinning parameters deduced from 50 velocity curves reported in the literature for different material and temperature and then proposes a model, which relates those parameters to microscopic properties of pinning.
A comparison of microscopic parameters characterizing the pinning and an analysis of 
thermal effects is presented in Section~\ref{sec:DWscales}.
In Section~\ref{sec:conclusions} we overview our results and summarize our main conclusions.

\section{Domain wall dynamics}
\label{sec:DWdynamics}

After a qualitative description of different magnetic field driven DW dynamical regimes observed experimentally, a self-consistent empirical approach exploiting the universal features of the creep and depinning regimes, is presented.
In this way we obtain the three fundamental pinning parameters which we use in the next Section to compare different magnetic materials.

\subsection{Different dynamical regimes}

\begin{figure}
\includegraphics[width=0.5\textwidth]{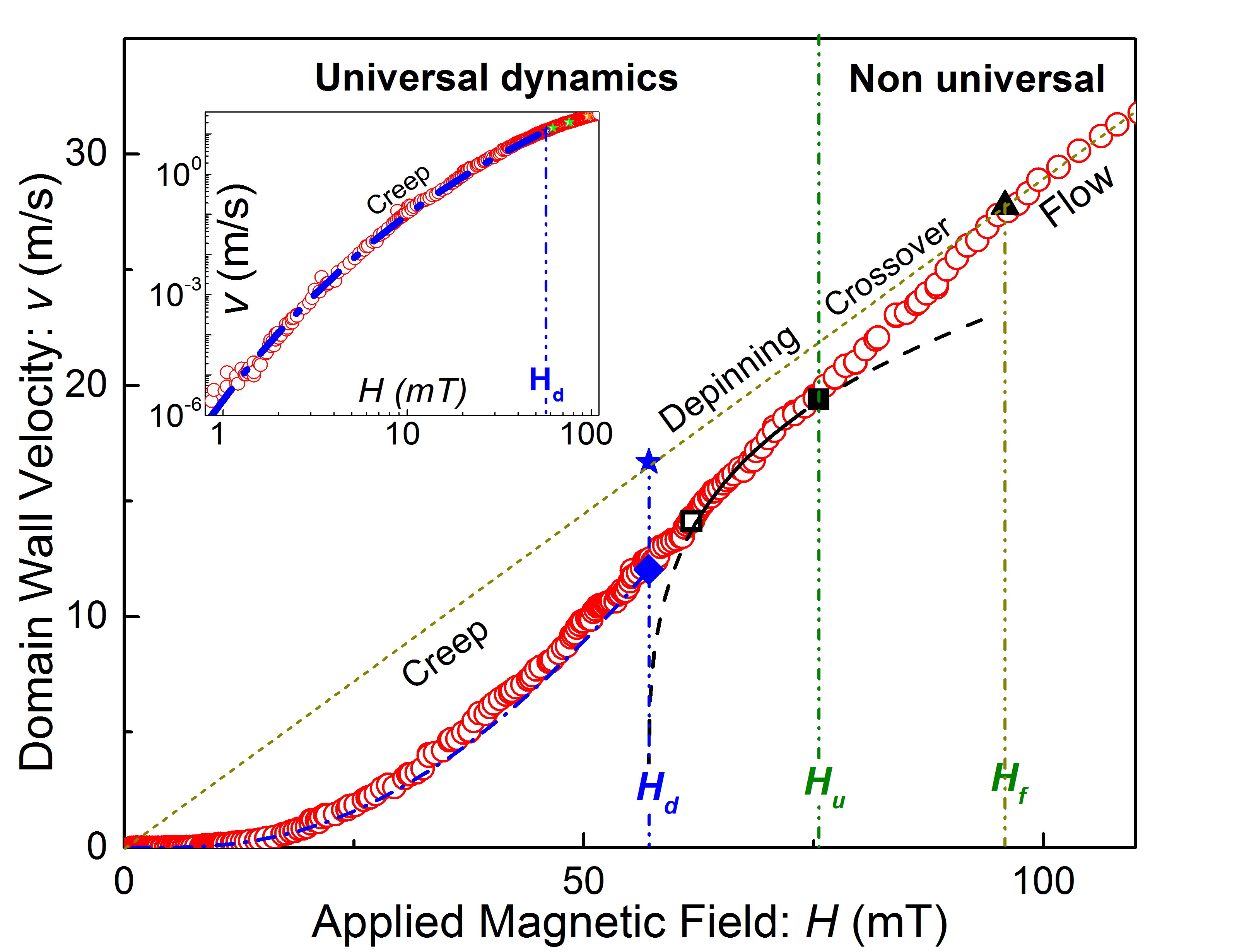}
\caption{\label{Fig1} Typical velocity curve observed for an ultrathin Pt/Co/Pt film at room temperature taken from Ref.~\onlinecite{diaz_PRB_2017_depinning}. The universal and the non-universal dynamics are separated by the boundary field $H_u$, which corresponds to the upper limit of the depinning transition. Within the non-universal dynamics, DWs present a crossover between the depinning transition and the linear flow regime, which is observed at the largest drive.
Within the universal dynamics ($H<H_u$), the depinning threshold $H=H_d$ separates the creep regime ($H<H_d$) from the depinning transition ($H_d<H<H_u$). Inset: log-log plot of the velocity curve highlighting the creep regime. 
The dotted line in the main panel corresponds to the linear extrapolation of the flow regime, observed for $H>H_f$. The dash dot curve is a fit of Eqs. \ref{eq:v-creep}, \ref{eq:E-creep} for the creep regime. The dashed curve is a fit of Eq. \ref{eq:depinning_H} for the depinning transition using the universal parameter $x_0 = v_T/v_H = 0.65$. The part that matches with experimental data is underlined by black solid segments. The diamond and star points located on the vertical line $H=H_d$ are the velocity at depinning $v(H_d)$ and the depinning velocity $v_T$. $v(H_d)$ corresponds to the inflection point separating the creep regime from the depinning transition. The value of $v_T$ was deduced from Eq. \ref{eq:depinning_T} and is found to coincide with the flow velocity DWs would have in absence of pinning.}
\end{figure}

A typical velocity curve of domain wall obtained for a Pt/Co/Pt ultrathin film is shown in Fig. \ref{Fig1} and is used to describe the different dynamical regimes. 
At low drive ($H<H_d$), the DWs move in the creep regime which is controlled by pinning, DW elasticity and thermal activation. The DW velocity follows an Arrhenius law $v \sim \exp(-\Delta E/k_BT)$, where $k_BT$ is the thermal activation energy, and $\Delta E$ the effective pinning barrier height.  The creep regime presents a universal behavior. Close to zero drive ($H \rightarrow 0$), the barrier height follows a power law variation with magnetic field $\Delta E \sim H^{-\mu}$ where $\mu$ is the so-called creep exponent\cite{lemerle_PRL_1998_domainwall_creep,chauve2000}. Increasing the applied magnetic field reduces the effective barrier height which vanishes ($\Delta E \rightarrow 0$) at the depinning threshold ($H=H_d$)\cite{gorchon_PRL_2014,jeudy_PRL_2016_energy_barrier}. 
Above the threshold, the curvature of velocity curve becomes negative ($d^2v/dH^2<0$)~\cite{diaz_PRB_2017_depinning}  (see Fig. \ref{Fig1}). The DWs undergo a depinning transition controlled by elasticity and thermal noise, which is also 
a universal dynamical regime~\cite{diaz_PRB_2017_depinning}. At zero temperature, the velocity is expected to follow a power law scaling with magnetic field $v \sim (H-H_d)^\beta$, where $\beta$ is the so-called depinning exponent. At finite temperature, the thermal activation produces a "thermal rounding" of the velocity curve.
The velocity is also predicted to present an asymptotic power law scaling with temperature at the depinning threshold $v \sim T^\psi$, where $\psi$ is the thermal rounding exponent~\cite{middleton_PRB_92_CDW_thermal_exponent, Roters_PRE_1999_depinning, bustingorry_epl_2008_thermal_rounding, bustingorry_PRB_12_thermal_rounding, bustingorry_PRE_12_thermal_rounding,diaz_PRB_2017_depinning}. 
The end of depinning transition corresponds to the onset of the divergence between the velocity curve and the magnetic field scaling law which crossovers at a magnetic field $H \simeq H_u$ (see Fig.~\ref{Fig1}). This upper boundary also roughly defines the limit of universal dynamics which covers the whole creep regime and the depinning transition~\cite{jeudy_PRL_2016_energy_barrier,diaz_PRB_2017_depinning}. 
Below $H=H_u$, the dynamics of a DW can be described as the motion of an elastic string submitted to thermal activation and to pinning\cite{chauve2000,ledoussal_PRB_2002_depinning_transition} and is independent of its magnetic structure. The measured critical exponents for a DW moving in an ultrathin film are compatible with theoretical predictions ($\mu=1/4$~\cite{chauve2000, lemerle_PRL_1998_domainwall_creep,metaxas_PRL_07_depinning_thermal_rounding,jeudy_PRL_2016_energy_barrier}, $\beta=0.25$~\cite{Ferrero_PRE_2013_depinning}, and $\psi=0.15$\cite{bustingorry_epl_2008_thermal_rounding,bustingorry_PRE_12_thermal_rounding,gorchon_PRL_2014,diaz_PRB_2017_depinning}) for the quenched Edwards Wilkinson universality class, with random short range uncorrelated pinning disorder.

%
Above $H=H_u$, the DW dynamics is
found to be of non-universal nature. The DW presents a crossover from the depinning transition to a flow regime. 
In the flow regime, the velocity depends on the time evolution of DW magnetic texture and presents a non-monotonous variation with magnetic field. Below the Walker limit~\cite{malozemoff} $H\leq H_w=(1/2)\alpha M_s$, where $\alpha$ is the so-called Gilbert damping parameter and $M_s$ the saturation magnetization, DW is expected to follow the so-called steady state regime for which its magnetic texture remains fixed during the motion. Above $H_w$ the DW velocity presents a negative slope~\cite{malozemoff} with the drive and it recovers a linear asymptotic variation at sufficiently high drive ($H \gg H_w$) which corresponds to the so-called asymptotic precessional regime.
 Experimentally, the steady state regime is rarely observed~\cite{dourlat_prb_2008, Beach_NatMat2005_V_H, thevenard_prb_2011}. As the Walker field is much smaller than the depinning field ($H_w \ll H_d$), it is generally hidden by pinning~\cite{metaxas_PRL_07_depinning_thermal_rounding}. This is the case for the curve of Fig. \ref{Fig1} where the linear variation corresponds to the precessional asymptotic regime~\cite{diaz_PRB_2017_depinning,metaxas_PRL_07_depinning_thermal_rounding}. 

%


\subsection{Universal glassy dynamics}
\label{sub:model_univ}
As shown in Refs.~\onlinecite{jeudy_PRL_2016_energy_barrier,diaz_PRB_2017_depinning},
the universal features of driven glassy DW dynamics, including the whole creep regime and the depinning transition, can be made explicit by introducing the reduced variables $H/H_d$, $T/T_d$, and $v/v_T$, where $H_d$, $T_d$, and $v_T$ are material and temperature dependent parameters characterizing DW pinning.
It is worth stressing that such description is self-consistent: the velocity of depinning and creep regimes are described by universal (though very different) functions of the \textit{same set} of three above-mentioned reduced variables. In the following, we describe the form of such functions. Table~\ref{table:table2} presents an overview of parameters describing the DW dynamics.

%

For the creep regime $[0<H<H_d(T)]$, the DW velocity is described by an Arrhenius law:   
\begin{equation}
 v(H,T) = v(H_d,T) \exp\left(-\frac{\Delta E}{k_B T} \right)
 \label{eq:v-creep}
\end{equation}
with the effective pinning barrier height given by 
\begin{equation}
 \Delta E= k_B T_d(T) \left[ \left( \frac{H}{H_d} \right)^{-\mu} - 1 \right],
 \label{eq:E-creep}
\end{equation}
where $k_B T_d$ is the characteristic pinning energy scale and $\mu$ ($=1/4$) the universal creep exponent. $v(H_d,T)$ corresponds to the velocity at depinning. In Ref. \onlinecite{jeudy_PRL_2016_energy_barrier}, it was shown that the ratio $\Delta E/k_B T_d$ is a universal function of the reduced magnetic field $H/H_d$, (i.e., material and temperature independent) which controls the creep velocity in whole $0<H<H_d$ range. 
The asymptotic behaviors of the pinning barrier height are a power law divergence $\Delta E \sim (H/H_d)^{-\mu}$ close to zero drive ($H \rightarrow 0$) and a linear collapse $\Delta E \sim \mu (1-H/H_d)$ close to the depinning threshold ($H \rightarrow H_d$). 

For the depinning transition $[H_d(T)<H<H_u(T)]$, the combined contributions of magnetic field and temperature on the velocity are described by a generalized universal homogeneous function~\cite{stanley_book_1971, privman_book_1991, Roters_PRE_1999_depinning, bustingorry_PRE_12_thermal_rounding} of the form:
\begin{equation}
y= g\left(\frac{x}{x_0} \right),
\label{eq:g_function}
\end{equation}
where the scaled dimensionless variables are defined as $y=(v/v_T)(T/T_d)^{-\psi}$ and $x=[(H-H_d)/H_d]^\beta (T/T_d)^{-\psi}$. A rather good approximation for the shape~\cite{diaz_PRB_2017_depinning} of the $g$-function is 
\begin{equation}
g(x/x_0)=\left[1+\left(x/x_0\right)^n \right]^{1/n},
\label{eq:g_function_shape}
\end{equation}
where $n$ ($=8.7 \pm 0.4$) tunes the width of the crossover and $x_0=0.65 \pm 0.04$ is a universal constant.
%
The DW velocity presents two universal asymptotic power law behaviors. At the depinning threshold ($H=H_d$), the temperature variation can be written as
\begin{equation}
\label{eq:depinning_T}
 v(H_d,T) =  v_T(H_d,T) \left( \frac{T}{T_d} \right)^{\psi},
\end{equation}
where $\psi$ ($=0.15$) is a depinning exponent and $v_T(H_d,T)$ a depinning velocity. 
%
Just above the depinning threshold~\cite{diaz_PRB_2017_depinning}, for $H \gtrsim H_d[1+(0.8(T_d/T)^{-\psi})^{1/\beta }]$, 
 the velocity is dominated by the driving field as
\begin{equation}
v(H,T)\approx \frac{v_T(H_d,T)}{x_0} \left( \frac{H-H_d}{H_d}\right)^{\beta},
\label{eq:depinning_H}
\end{equation}
where $\beta$ ($= 0.25$) is another depinning exponent. For most of the studied magnetic materials, the thermal activation energy is much smaller than the pinning energy ($T \ll T_d$), and part of the velocity curve just above $H_d$ present a good agreement with the predictions of Eq. \ref{eq:depinning_H} as shown in Fig. \ref{Fig1}. 

In summary, the set of Eqs. \ref{eq:v-creep}, \ref{eq:E-creep}, and \ref{eq:g_function} constitutes a self-consistent description of the DW glassy dynamics observed below the universality limit ($H \leq H_u$). The creep motion and the depinning transition are both described by universal functions and their asymptotic limits agree with the predictions from models of elastic lines in disordered media. 
The non-universal character of DW motion is caught by only three purely material and temperature dependent parameters corresponding to the depinning threshold $H_d$, temperature $T_d$, and velocity $v_T$. 

\subsection{Self consistent analysis of DW dynamics}

The determination of material and temperature dependent parameters requires to perform simultaneously a fit of the creep regime (Eqs. \ref{eq:v-creep} and \ref{eq:E-creep} ) with adjustable velocity-magnetic field coordinates at depinning ($H_d$, $v(H_d)$), and of the depinning transition (Eq. \ref{eq:depinning_H}) over an adjustable range (with an upper bound $H=H_u$). 
The following procedure can be used: 
\begin{itemize}
	\item Step 1: the upper boundary of the creep regime ($H_d$, $v(H_d)$) is assumed to correspond to the inflection point of the velocity curves (see the diamond in Fig. \ref{Fig1}). Indeed, the curvature is predicted to change of sign at the depinning transition: positive for the creep regime ($H<H_d$, see Eqs. \ref{eq:v-creep} and \ref{eq:E-creep}) and negative for the depinning regime ($H>H_d$, see Eqs. \ref{eq:depinning_H}).
	\item Step 2: an estimate of $T_d$ is then deduced from a fit of $v(H)$ with Eqs. \ref{eq:v-creep} and \ref{eq:E-creep} (with $\mu=1/4$) over the range $0<H<H_d$ (see the dot-dash line in Fig. \ref{Fig1}).
	\item Step 3: in order to improve the accuracy for the values of $H_d$ and $v(H_d)$ a fit of Eqs. \ref{eq:v-creep} and \ref{eq:E-creep} is performed for increasing values of $H$. The upper boundary of the creep regime ($H_d$, $v(H_d)$) can also be defined as the limit above which the fit and the experimental curve start to diverge. The step 2 can then be repeated to improve the accuracy for the $T_d$-value.
	\item Step 4: a final fine tuning of $H_d$ and $v(H_d)$ is deduced from a fit of Eq. \ref{eq:depinning_H} with $\beta=0.25$ and $x_0=0.65$ over the largest magnetic field range (see the dashed curve in Fig. \ref{Fig1}).
	\item Step 5: when the linear flow regime is observed (as in the case of Pt/Co/Pt films), the coordinates ($H_d$, $v(H_d)$) can be also finely adjusted using Eq. \ref{eq:depinning_T} and assuming $v_T$ to coincide with the velocity of linear flow regime~\cite{diaz_PRB_2017_depinning} (see the star in Fig. \ref{Fig1}) . 
\end{itemize}
This procedure was used to analyze 50 velocity curves reported in the literature. 

\section{Domain wall pinning}
\label{sec:DWpinning}

In this section, we first present the effective pinning parameters ($H_d$, $v_T$, and $T_d$) deduced from the analysis of the glassy dynamics for different materials and various temperatures.
We then propose a model, which relates those parameters to the micromagnetic and microscopic pinning parameters.  

\subsection{Effective pinning parameters}

\begin{figure}
\includegraphics[width=0.48\textwidth]{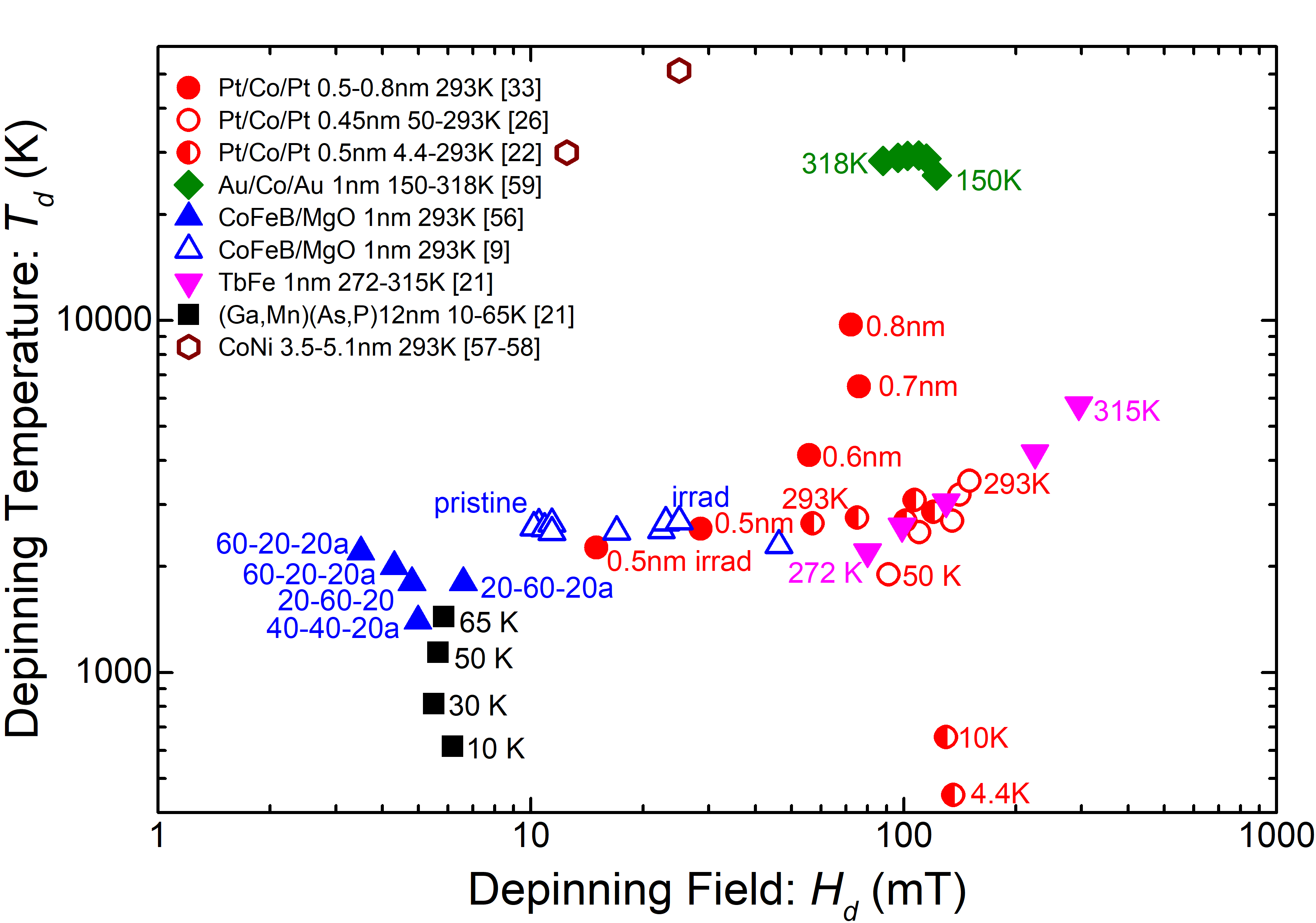}
\caption{\label{fig:Td_Hd} Depinning temperature $T_d$ versus depinning field $H_d$. For each magnetic material, the legend indicates the film thickness, the explored temperature range, and the reference. For the CoFeB films, the letter “a” means annealed. 
}
\end{figure}

\begin{figure}
	\includegraphics[width=0.48\textwidth]{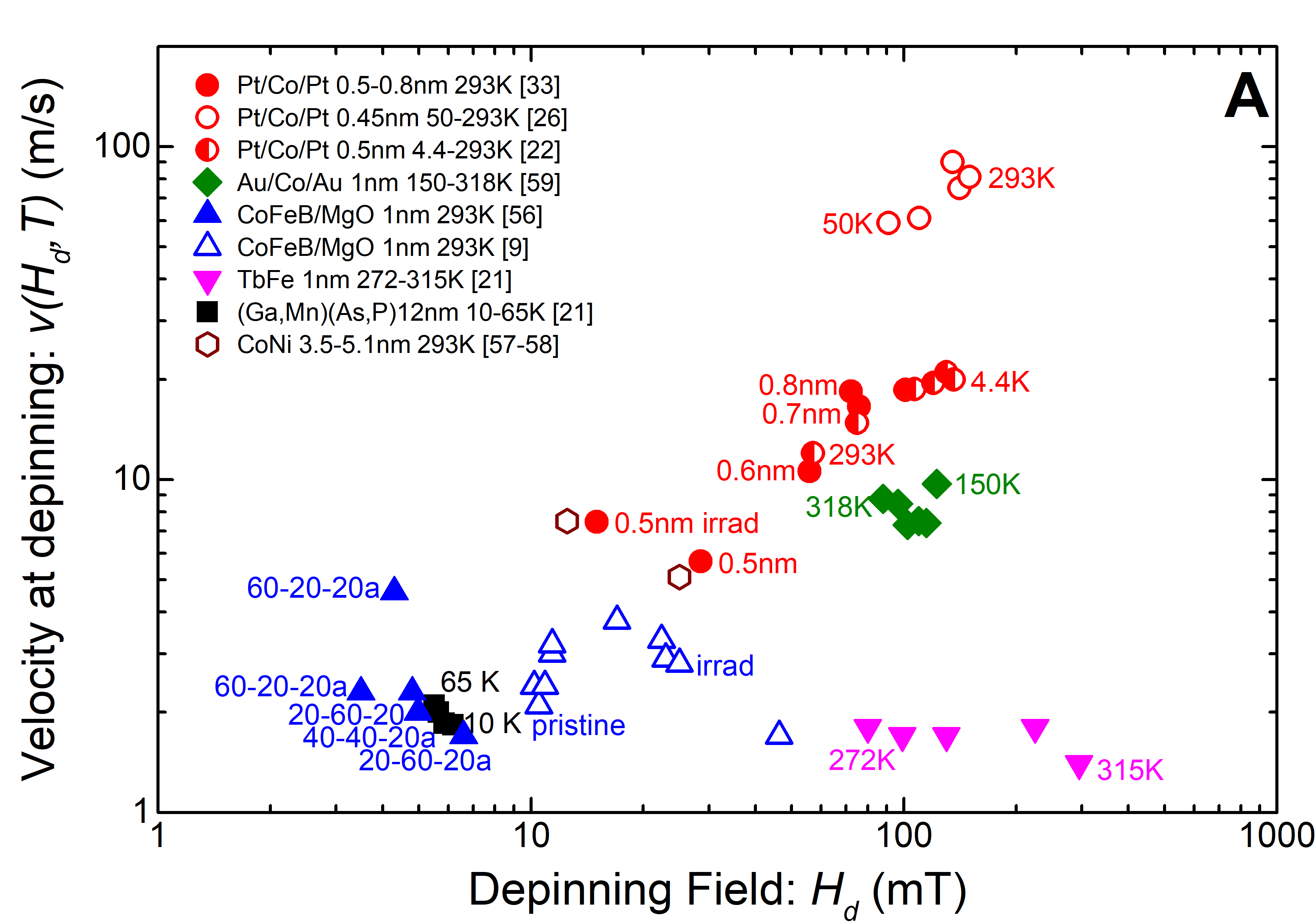}
	\includegraphics[width=0.48\textwidth]{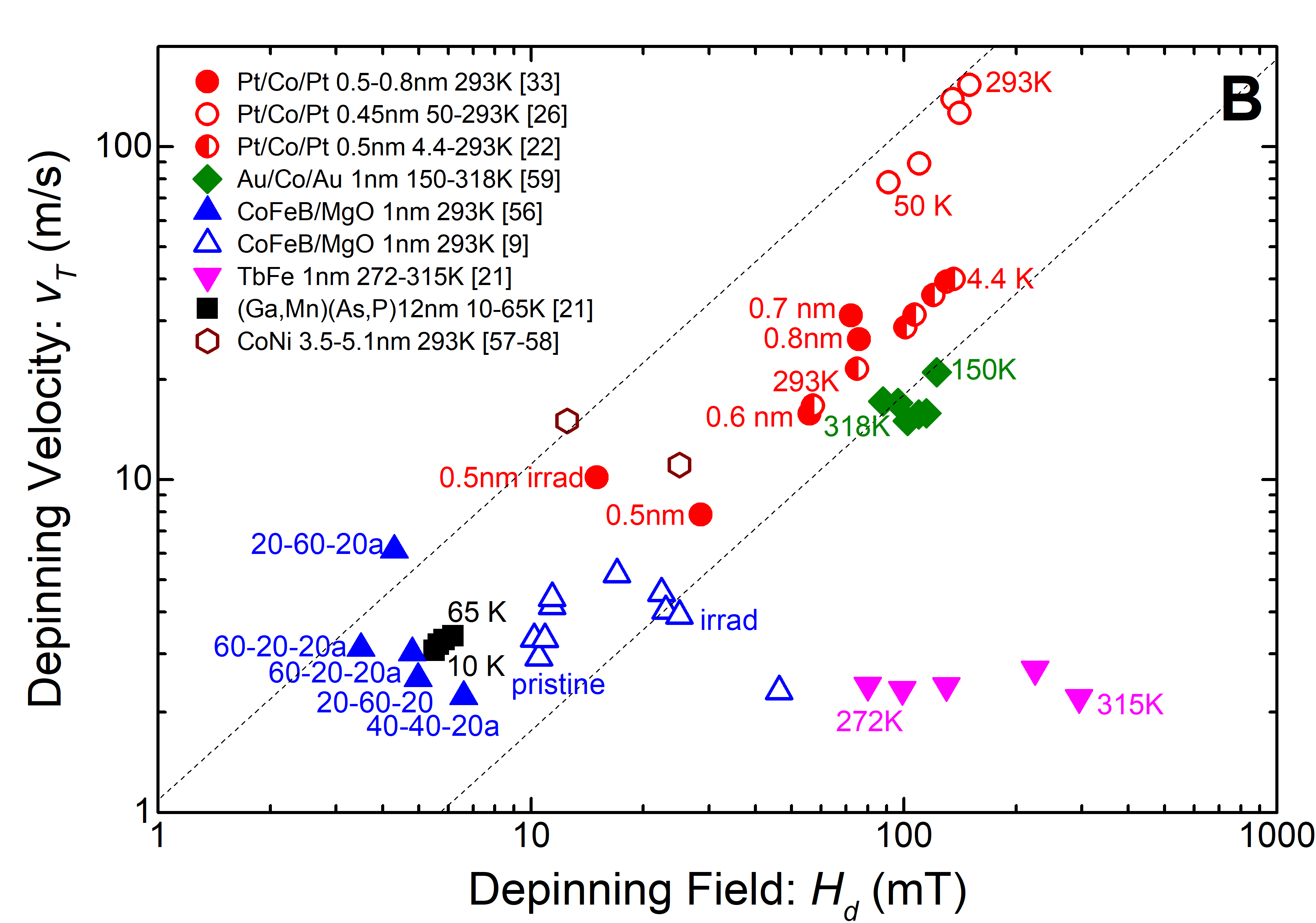}
	\caption{\label{fig:v(Hd)_Hd} (A) Velocity at depinning $v(H_d)$ and (B) depinning velocity $v_T$ versus depinning field $H_d$. $v(H_d)$ corresponds to the measured upper limit of the creep velocity. $v_T(H_d)$ was deduced from Eq. \ref{eq:depinning_T} and is only dependent on material and temperature. The global increasing trend of $v_T(H_d)$ is materialized by two dash lines which correspond to linear variations of depinning velocity ($v_T=mH_d$) for two constant slopes ($m=0.15$ and $1.1$~m/(s.mT)). For the CoFeB films, the letter “a” means annealed.  
	}
\end{figure}

A synoptic presentation of the effective pinning parameters is proposed in Figs.~\ref{fig:Td_Hd}, and \ref{fig:v(Hd)_Hd}. See also the Table \ref{table:table1} in the annex for details and for the values of micromagnetic parameters.  

A plot of the depinning field $H_d$ versus depinning temperature $T_d$ is shown in Fig.~\ref{fig:Td_Hd}.  
As it can be observed, the data points are rather dispersed. The values of $H_d$ and $T_d$ extend over two orders of magnitude ($H_d$: from $3mT$ for CoFeB/MgO to $300mT$ for TbFe, and $T_d$: from $600K$ for (Ga,Mn)(As,P) to $50000K$ for CoNi). From Fig.~\ref{fig:Td_Hd},it is not evident to extract general trends for the variations of the effective height pinning barrier $k_BT_d$ with the depinning threshold $H_d$. The analysis of those variations is extensively discussed in the following.

Now let us discuss the results obtained for the velocity at depinning $v(H_d)$ 
and the depinning velocity $v_T(H_d)$ (see Eq. \ref{eq:depinning_T}).
As shown in Fig. \ref{fig:v(Hd)_Hd}~A, $v(H_d)$ globally increases with increasing $H_d$-value and covers a typical range extending from $1$ to $100$~m/s. 
The significantly lower values obtained for TbFe compared to others magnetic materials could be due to an underestimation of the depinning threshold since no change of curvature sign was observed in the velocity curve of Ref. \onlinecite{jeudy_PRL_2016_energy_barrier}.
For the depinning velocity $v_T$, a global increase with $H_d$ is observed in Fig. \ref{fig:v(Hd)_Hd} B. This trend can be framed by two linear variations of the depinning velocity ($v_T=mH_d$) for two constant slopes (see the dashed lines which correspond to $m=0.15$ and $1.1$m/(s.mT)). 

More generally, the map of effective pinning parameters presented in Figs. \ref{fig:Td_Hd} and \ref{fig:v(Hd)_Hd} B is a main result of the present paper. It serves as a starting point for the discussion on the microscopic origin of pinning proposed in the following.

\subsection{Model for domain wall pinning}
\label{sub:model_depinning}

In order to go further in the analysis of pinning properties, the effective pinning parameters deduced from velocity curves ($H_d$, $v_T$, and $T_d$) have to be related the microscopic characteristics of pinning and to ultimately to the micromagnetic parameters of each material. 

The universal functions 
describing the velocity are
consistent with the general theoretical predictions obtained by solving the large-scale non-equilibrium behavior of a driven elastic string in random medium~\cite{Ferre_CRP_2013_review}.
We thus expect to obtain $H_d$, $v_T$, and $T_d$ from a simple dimensional analysis of such  model at relatively short length-scales. 
We follow the approaches
of Refs.~\onlinecite{nattermann_prb_1990,chen_prb_1995,lemerle_PRL_1998_domainwall_creep} which consider the DW as an elastic line, not taking into account the detailed magnetic 
texture ~\cite{lecomte_PRB_2009_internal_degree_freedom,moretti_PRB_17} and express the model parameters in terms of micromagnetic quantities.

The variation of the free energy associated with the displacement of a DW segment of length $L$ over a transverse distance $u$ is roughly given by:
\begin{equation}
\label{eq:free_energie}
\delta F(L,u)=\sigma t u^2 / L + \delta F_{pin} (L,u)-2M_s H t L u,
\end{equation}
where the magnetization saturation $M_s$, the DW elastic energy $\sigma$ ($\approx 4 \sqrt{AK_{eff}}$ for a Bloch wall, where $A$ and $K_{eff}$ are the stiffness and the effective anisotropy constant, respectively) are micromagnetic parameters, and $t$ is the layer thickness (see Table~\ref{table:table2} for an overview of the parameters controlling domain wall pinning).
In Eq.~(\ref{eq:free_energie}), the first term corresponds to the elastic energy associated to the elongation of domain wall, $\delta F_{el}$, the second is the pinning energy, $\delta F_{pin}$, and the third term stands for the contribution of the driving magnetic field, $\delta F_H$. 

The DW is expected to be depinned for an applied magnetic field $H=H_d$, such that no metastable states with zero velocity exist for $H>H_d$. 
Larkin realized that this happens when a DW segment of a characteristic size $L_c$, displaces over the characteristic range of effective pinning potential ($u\approx \xi$), or pinning force correlation length, in response to the field. The so-called Larkin length $L_c$ is field independent and can be estimated from 
$\delta F_{pin}(\xi,L_c) \approx \delta F_{el}(\xi,L_c)$. This estimate can be explicitly done after modeling the scaling properties of $\delta F_{pin}(u,L)$, typically using collective pinning theory if pinning is weak~\cite{larkin1979pinning,blatter_vortex_review}.

%
At depinning, the elastic 
and Zeeman terms are of the same order ($\delta F_{el}(L_c,\xi)\approx \delta F_H(L_c,\xi)$), 
establishing a connection between the collectively pinned DW segment length $L_c$ and the depinning field $H_d$. 
Moreover, we can assume that the typical 
pinning energy barrier height encountered in Eqs. \ref{eq:E-creep} and \ref{eq:depinning_T} corresponds to the pinning energy at the depinning threshold $k_B T_d \approx \delta F_{pin} (L_c,\xi)$. As the latter energy contribution should be also of the same order as $\delta F_{el}(L_c,\xi)$, using Eq.~(\ref{eq:free_energie}) we obtain  
  
%
\begin{eqnarray}
\label{eq:relation1}
k_B& T_d=&(\xi^2\sigma t )/ L_c  \\
\label{eq:relation2}
H_d&=&\sigma \xi/(2 M_s L_c^{2}),
\end{eqnarray}
 which relate the depinning temperature and magnetic field to the microscopic length scales $L_c$ and $\xi$, and to the micromagnetic parameters $\sigma$ and $M_s$. 
 %
 As this model is essentially based on scaling arguments, it is expected to describe 
 correctly the temperature variation of $L_c$ and $\xi$ and to provide a rough estimate of their magnitude from the knowledge of velocity response parameters $T_d$ and $H_d$.
 
 
%
Let us now discuss the DW velocity at depinning $v(H_d,T)$ and the depinning velocity $v_T$. 
In Eqs. \ref{eq:depinning_T} and \ref{eq:depinning_H}, $v_T$ is defined as a purely scaling factor and it is important to give to this parameter a precise physical meaning. Following the discussions in Refs.~\onlinecite{gorchon_PRL_2014,diaz_PRB_2017_depinning}, we assume that $v_T$ corresponds to the velocity, which would have DWs in the absence of pinning, which yields
\begin{equation}
\label{eq:depinning_velocity}
v_T(H_d,T)=m_{fl}(H_d,T)H_d(T),
\end{equation}
where $m_{fl}(H_d,T)$ is the mobility of the DW in the flow regime at the depinning field $H_d$. Strictly, $m_{fl}(H_d,T)$ has a non-monotonous field dependence with two important reference values: $m_{fl}(H_d,T)=\gamma \Delta /\alpha$ for $H_d \leq H_w$ and 
\begin{equation}
\label{eq:mobility}
m_{fl}(H_d,T)=\frac{\gamma \Delta\alpha}{1+\alpha^2},
\end{equation}
for $H_d \gg H_w$, where $\Delta=\sqrt{A/K_{eff}}$ is the domain wall width parameter, $\alpha$ the Gilbert damping parameter, and $\gamma$ the gyromagnetic factor ($=1.76 10^{11}$~Hz.T$^{-1}$).
For the data of Refs.\onlinecite{metaxas_PRL_07_depinning_thermal_rounding,diaz_PRB_2017_depinning,jeudy_PRL_2016_energy_barrier}, as in the case of Fig. \ref{Fig1}, $m_{fl}(H_d,T)$ corresponds to the mobility of asymptotic precessional flow regime (i.e., $H_w \ll H_d$).
%


\subsection{Thermal effects}

The thermal activation produces fluctuation of DW position which, if strong enough, can appreciably smooth the effective random pinning potential experienced by DWs. As a result, the correlation length of the disorder $\xi(T)$ and the Larkin length $L_c(T)$ are expected to increase with increasing temperature. For a weakly pinned elastic line, Nattermann {\it et al.}~\cite{nattermann_prb_1990,chen_prb_1995} proposed the following temperature variations: $\xi(T)=\xi_0[1+(T/T_d)]^{3}$ and $L_c(T)=L_{c0}[1+(T/T_d)]^{5}$, which are interpolation formula between power law variations deduced form scaling arguments and values ($\xi_0$ and $L_{c0}$) corresponding to the limit of zero thermal fluctuation of DW position. More recently, Agoritsas {\it et al.} \cite{agoritsas_PRE_13} proposed the analytic predictions:
 \begin{equation}
\label{eq:larkin_agoritsas}
L_c(T,\xi)=\frac{4\pi}{\sigma t D^2} \left(\frac{k_BT}{f(T/T_d)}\right)^5
\end{equation} 
 and
 \begin{equation}
\label{eq:ksi_agoritsas}
\xi(T)=\frac{\sqrt{3} (4\pi)^{5/6}}{\sigma t D} \left(\frac{k_BT}{f(T/T_d)}\right)^3,
\end{equation}
where the function $f$ is given by the implicit equation $f^6=4\pi (1-f)(T/T_d)^6$.
The zero temperature values are given by:  
\begin{equation}
\label{eq:larkin0_agoritsas}
L_{c0}=(4\pi)^{1/6}\frac{(k_BT_d)^5}{\sigma t D^2}
\end{equation} 
 and
 \begin{equation}
\label{eq:ksi0_agoritsas}
\xi_0=\sqrt{3} (4\pi)^{5/6}\frac{(k_BT_d)^3}{\sigma t D}.
\end{equation} 
%
In Eqs. \ref{eq:larkin_agoritsas}, \ref{eq:ksi_agoritsas}, \ref{eq:larkin0_agoritsas}, and \ref{eq:ksi0_agoritsas}, $D$ is the strength of disorder~\cite{kolton_prb_2009_pathways,agoritsas_physica_B_2012,agoritsas_PRE_13},
 reflecting the typical amplitude of the quenched random pinning potential and has the dimension of the square of an energy. Note that Eqs. \ref{eq:larkin0_agoritsas} and \ref{eq:ksi0_agoritsas} indicate that $L_{c0}$ and $\xi_0$ are also expected to present intrinsic temperature variations due to their dependency on micromagnetic and pinning parameters. Such intrinsic variation must be hence distinguished from the extrinsic variation explicitly given in Eqs.~\ref{eq:larkin_agoritsas} and \ref{eq:ksi_agoritsas}, which becomes important only when the temperature $T$ is close to $T_d$.


\subsection{Pinning and domain wall dynamics} 

In order of get a better insight into the variation of DW dynamics with the magnetic material and temperature, it is interesting to relate the pinning parameters (see Figs. \ref{fig:Td_Hd} \ref{fig:v(Hd)_Hd}) with the micromagnetic and microscopic pinning parameters.

A more intuitive insight of the predictions of Ref.~\onlinecite{agoritsas_PRE_13} can be deduced from scaling arguments. Following Ref. \onlinecite{lemerle_PRL_1998_domainwall_creep} and neglecting thermal effects (i.e. $L_{c}=L_{c0}$, and $\xi=\xi_0$) the pinning energy can be modeled by collective pinning theory \cite{larkin1979pinning} $\delta F_{pin} (L_{c0},\xi_0)=f_{pin}\xi_0 \sqrt{n \xi_0 L_{c0}}$, where $n$ is the density of pinning centers ($\approx 1/\xi_0^2$), and $f_{pin}$ a typical pinning force. Using Eqs. \ref{eq:relation1} and \ref{eq:relation2} leads to
\begin{equation}
\label{eq:kB_Td_2}
(k_BT_d)^3= \sigma t (f_{pin}\xi_0)^2 \xi_0
\end{equation} 
which for $f_{pin}^2\xi_0^2=D/(\sqrt{3} (4\pi)^{5/6})$ is equivalent to Eq. \ref{eq:ksi0_agoritsas} and to
\begin{equation}
\label{eq:Hd_2}
(H_d)^3=\frac{(f_{pin}\xi_0)^4}{\xi_0^7 \sigma t^4 (2M_s)^3},
\end{equation} 
respectively. Eqs. \ref{eq:kB_Td_2}, \ref{eq:Hd_2}, and  \ref{eq:depinning_velocity}, \ref{eq:mobility} now fully relate the pinning parameters controlling DW dynamics ($H_d$, $v_T$, and $T_d$)  to the micromagnetic parameters ($\sigma$, $M_s$, $\Delta$, and $\alpha$) and the microscopic pinning parameters ($f_{pin}$ and $\xi_0$). 
Therefore, combining the description of universal behaviors (see Eqs. \ref{eq:v-creep}, \ref{eq:E-creep}, \ref{eq:g_function}, \ref{eq:depinning_T}, and \ref{eq:depinning_H} in sect. \ref{sub:model_univ}), the predictions of model for DW pinning (see Eqs. \ref{eq:kB_Td_2} and \ref{eq:Hd_2}) and experimental measurements of the micromagnetic parameters, one can estimate the microscopic parameters controlling DW pinning. 

\section{Fundamental pinning scales
}
\label{sec:DWscales}
By mean of the scaling model of pinning developed in Sect. \ref{sec:DWpinning}, it is possible to discuss the fundamental pinning scales from the map (see Figs. \ref{fig:Td_Hd} and \ref{fig:v(Hd)_Hd} B) of material and temperature dependent pinning parameters controlling domain wall dynamics.

%

\subsection{Characteristic length-scales of pinning
}

Using Eqs. \ref{eq:relation1} and \ref{eq:relation2}, we can deduce the range of the pinning potential
\begin{equation}
\label{eq:xi}
\xi=[(k_B T_d)^2/(2 M_s H_d \sigma t^2)]^{1/3},
\end{equation}
and the Larkin length
\begin{equation}
\label{eq:Lc}
L_c=[(\sigma k_B T_d)/(4M_s^2 t H_d^2)]^{1/3}.
\end{equation}
Those relations are expected to provide estimations of values $\xi$ and $L_c$ and to reveal their temperature variation~\cite{gorchon_PRL_2014}.
Following Eqs. \ref{eq:xi} and \ref{eq:Lc}, estimations of $\xi$ and $L_c$ rely on the values of $M_s$ and $\sigma$. As it can be seen in Table~\ref{table:table1}, this eliminates the analysis for materials Au/Co/Au, CoFeB for different irradiation dose, and TbFe. 
Note that we could also consider the Larkin area $L_c \xi$ ($=(k_B T_d)/(2 H_d M_s t)$), which is independent of $\sigma$. 

\begin{figure}
\includegraphics[width=0.5\textwidth]{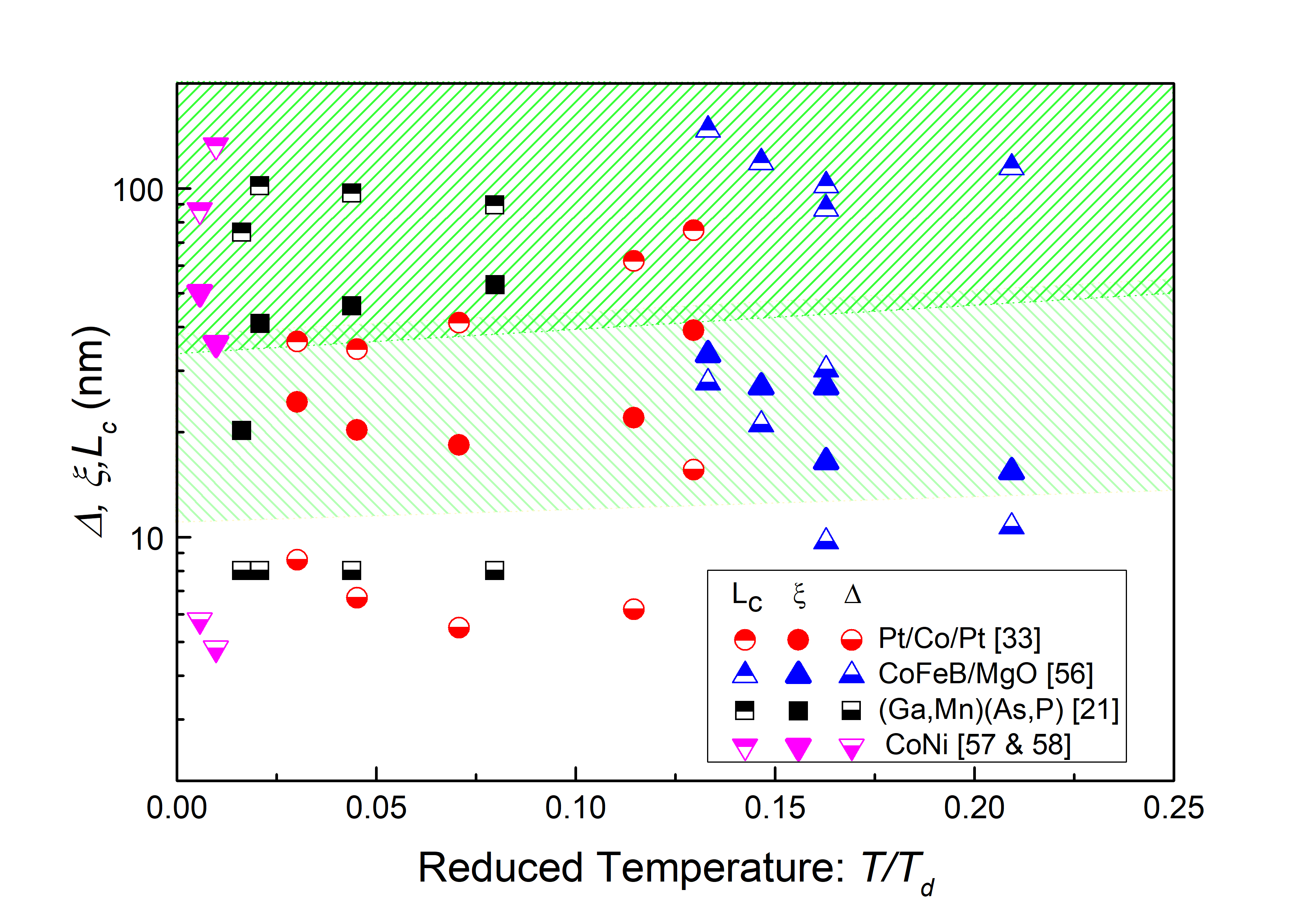}
\caption{\label{Fig: lengths} Characteristic lengths versus reduced temperature. $\Delta$, $\xi$, and $L_c$ are the domain wall width parameter, the correlation length of the disorder, and the Larkin length, respectively.}
\end{figure}

As it can be observed in Fig.~\ref{Fig: lengths}, for all the reported materials (Pt/Co/Pt, Co/Ni, (Ga,Mn)As, and CoFeB/MgO), the range of values for the pinning correlation length  ($\xi\approx 20-50nm$), and the Larkin length  ($L_c \approx 40-170nm$) are relatively well separated.
The ratio between $L_c$ and $\xi$ scales the density of pinning sites along the DWs. Its relatively small values suggest that DWs pinning involves only few pinning sites ($2-4$) over the Larkin length.
Moreover, the correlation length of the disorder is larger than the domain wall width parameter ($\Delta \approx 5-20nm$) except for CoFeB/MgO for which we have $\Delta \approx \xi$. This indicates that generally the weak pinning originates from fluctuations of pinning over distances larger than the domain wall width parameter.

\subsection{Temperature variations of the pinning strength and length-scales}

Let us now discuss the temperature variation of $\xi$ and $L_c$.
In order to compare the temperature variation for different materials and theoretical predictions we first normalize the data for each material to values of $\xi_n$ and $L_{cn}$, which are assumed to be temperature independent.
Those values were chosen in order for the ratios $\xi(T)/\xi_n$ and $L_c(T)/L_{cn}$ to follow the temperature variation predicted by Eqs.~\ref{eq:larkin_agoritsas} and \ref{eq:ksi_agoritsas} over the largest range of $T/T_d$, respectively.

\begin{figure}
\includegraphics[width=0.5\textwidth]{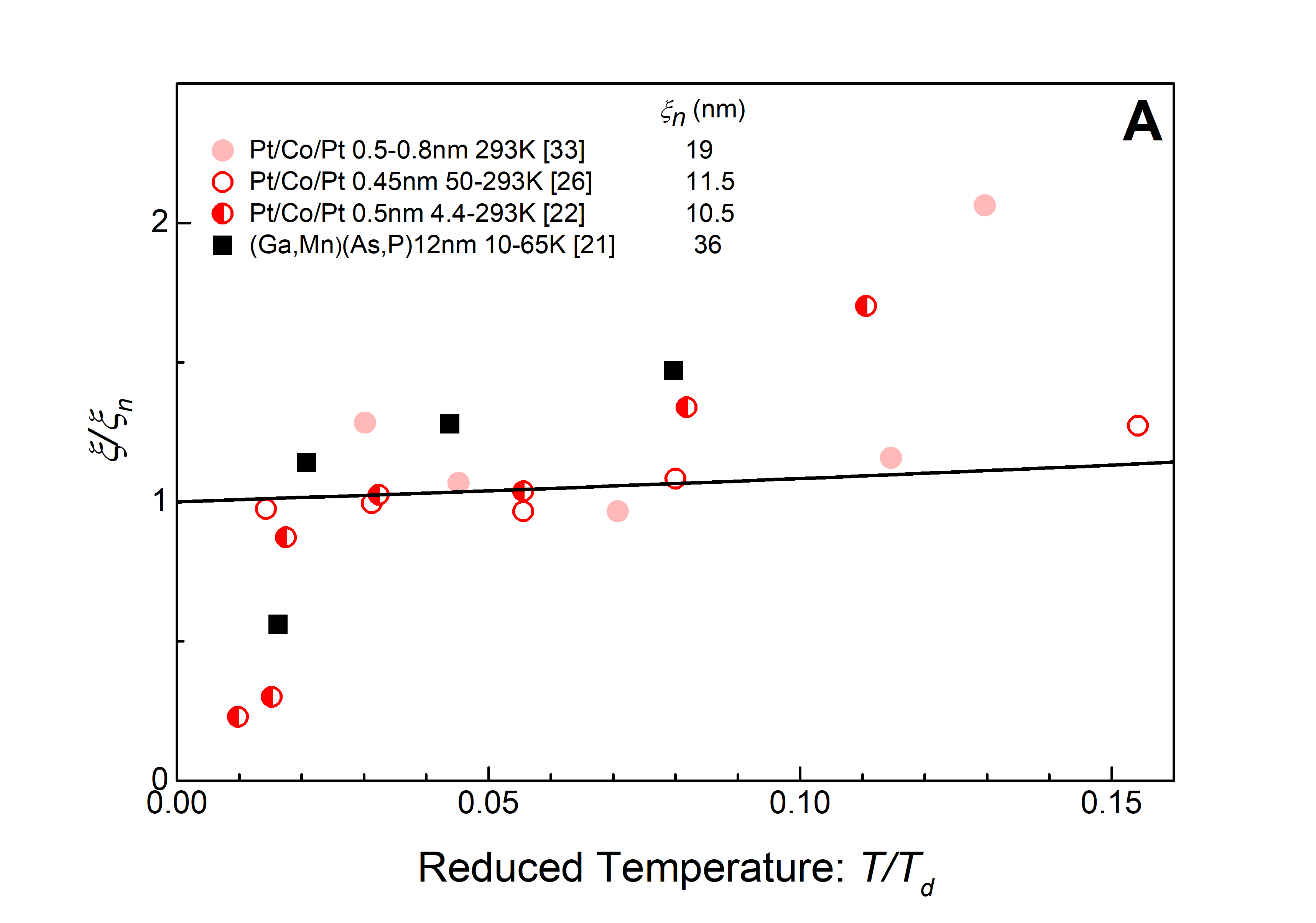}
\includegraphics[width=0.5\textwidth]{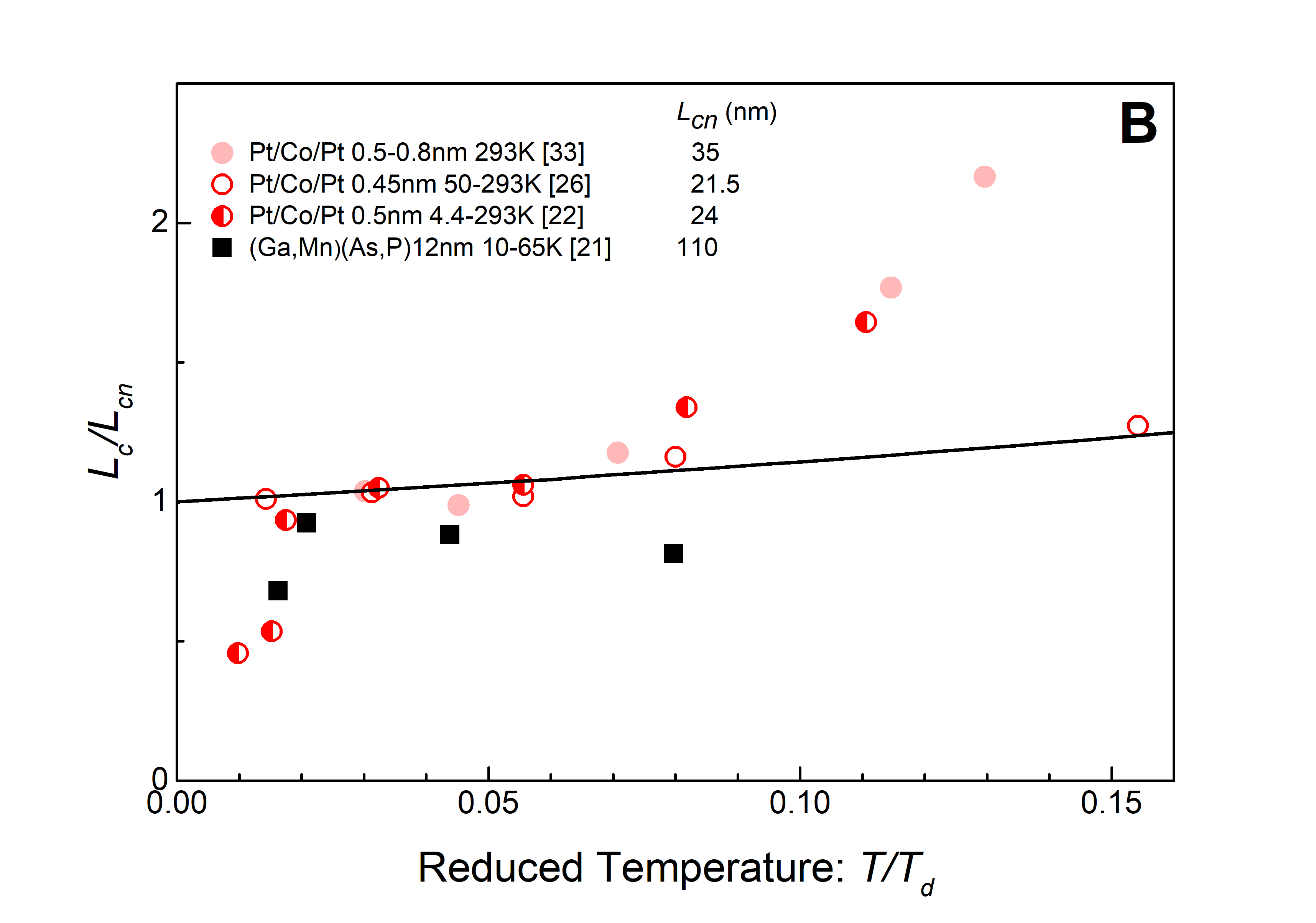}
\caption{\label{fig:thermal_effects} (A) Reduced correlation length of the disorder $\xi/\xi_n$, and (B) reduced Larkin length $L_c/L_{cn}$ as a function of reduced temperature $T/T_d$. The solid lines in A and B correspond to the predictions of Eqs. \ref{eq:larkin_agoritsas} and \ref{eq:ksi_agoritsas}, respectively. All the data correspond to single sample and variable temperature except the shade red points (Pt/Co/Pt: Ref.~\cite{metaxas_PRL_07_depinning_thermal_rounding}), which correspond to room temperature and different sample thicknesses. The normalization values ($\xi_n$ and $L_{cn}$) are indicated in the figures for each material. In the range $0.02<T/T_d<0.1$, the data  agree well with theoretical predictions. For $T/T_d<0.02$, there is a drop of both $\xi/\xi_n$ and $L_c/L_{cn}$.}
\end{figure}

The results are shown in Fig.~\ref{fig:thermal_effects} for materials for which both $\sigma$ and $M_s$ are reported in the literature. 
Following Eqs. \ref{eq:larkin_agoritsas} and \ref{eq:ksi_agoritsas},
$\xi/\xi_n$ and $L_c/L_{cn}$ are predicted to decrease rather weakly as the temperature is reduced. In contrast, the experimental data present an important variation with temperature. This suggests that the thermal behavior of $\xi(T)$ and $L_c(T)$ is dominated by the temperature variation of the micromagnetic and pinning parameters of $\xi_0(T)$ and $L_{c0}(T)$ (reflected by Eqs.~\ref{eq:larkin0_agoritsas} and \ref{eq:ksi0_agoritsas}) and not by thermal fluctuations of DW position.

Assuming now negligible fluctuations of DW position, the data of Fig.~\ref{fig:thermal_effects} can be viewed as the relative temperature variation of  $\xi_0(T)$ and $L_{c0}(T)$. 
Different regimes can be clearly distinguished. 
For $T/T_d>0.02$, $\xi_0(T)/\xi_n$ and $L_{c0}(T)/L_{cn}$ globally tend to weakly increase with temperature.
For $T/T_d>0.10$, the large observed fluctuations suggest a strong sample dependent temperature behavior. 
%
For $T/T_d<0.02$, both  $\xi_0(T)/\xi_n$ and $L_{c0}(T)/L_{cn}$ are observed to drop with decreasing temperature. The decrease of $\xi_0(T)/\xi_n$ can reach a factor 4, which suggests that $\xi_0(T)$ becomes close to the DW width $\Delta$. Therefore, we can infer the drop observed in Fig.~\ref{fig:thermal_effects} to reflect a crossover between different pinning length scales. At sufficiently large reduced temperature ($T/T_d>0.02$), the correlation length of the disorder is larger than the DW thickness ($\xi_0>\Delta$). At low temperature ($T/T_d<0.02$), the pinning is controlled by the DW width, which defines the correlation length of the disorder ($\xi_0 \approx \Delta$).

\begin{figure}
\includegraphics[width=0.5\textwidth]{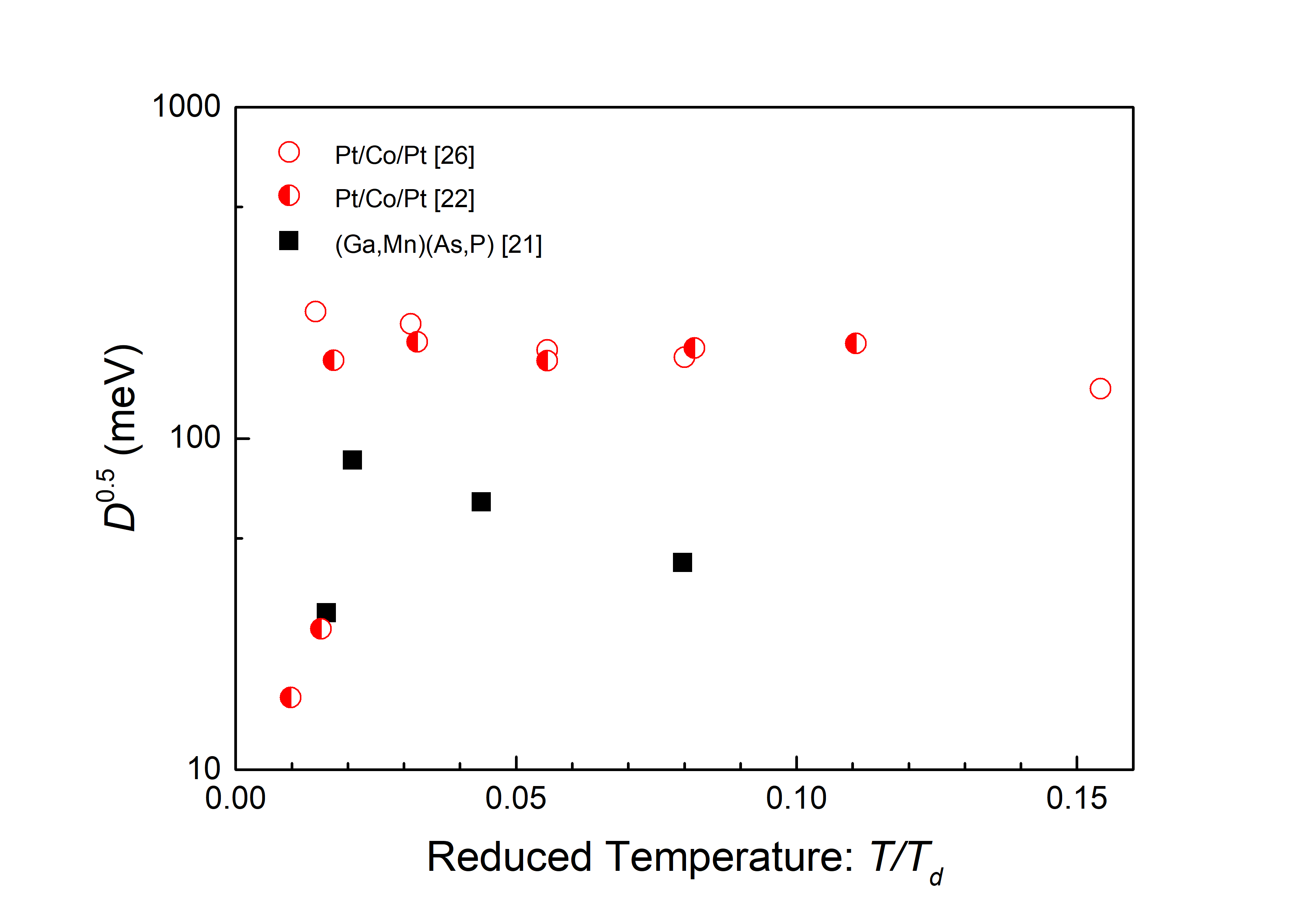}

\caption{\label{fig:pinning_disorder} Square root of the strength of the disorder as a function of reduced temperature $T/T_d$.}
\end{figure}

In order to further analyze the pinning, it is interesting to discuss the strength of pinning disorder. The value of the strength of pinning disorder can be deduced from Eq. \ref{eq:larkin0_agoritsas} ($D^2=(4\pi)^{1/6}[(k_BT_d)^5]/[\sigma t L_{c0}]$). In Fig. \ref{fig:pinning_disorder}, we plot $\sqrt{D}$ (which has the dimension of an energy) as a function of the ratio $T/T_d$. Above the crossover ($T/T_d>0.02$), the strength of pinning disorder is almost temperature independent for the Pt/Co/Pt films ($\sqrt{D} \approx 200$~meV), as expected for a quenched disorder. For (Ga,Mn)(As,P), the slight decrease of $\sqrt{D}$ with increasing temperature is probably associated to a not enough stringent estimation of the DW energy $\sigma$ (see Table \ref{table:table1}). 
For $T/T_d<0.02$, we observe a drop of the pinning strength. At low temperature, DWs become sensitive to pinning sites with both lower strength and shorter range, which are un-efficient at higher temperature. This also suggests the existence of different pinning strength ranges and a crossover between pinning regimes tuned by the magnitude of thermal activation. A better understanding of this issue requires further investigations. 

\subsection{Depinning velocity}

\begin{figure}
\includegraphics[width=0.5\textwidth]{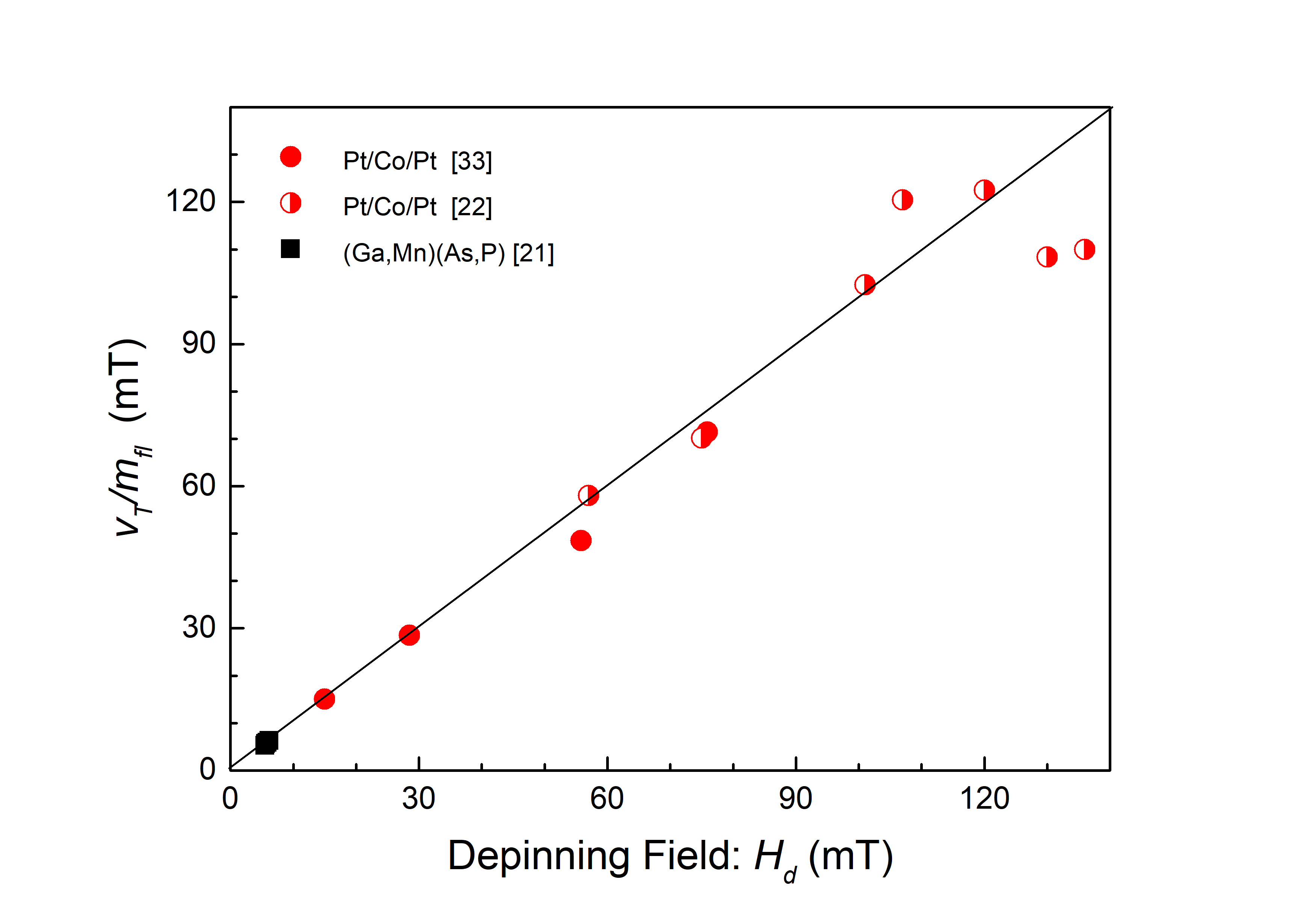}
\caption{\label{fig:R_v(Hd)_Hd} Ratio between depinning velocity and mobility in the flow regime versus depinning field. The equality between those two quantities indicates that the depinning velocity corresponds to the velocity DW would have in the absence of pinning.}
\end{figure}

Let us now discuss the depinning velocity $v_T$, which is important as it sets the fundamental time scale, once known the length-scales characterizing the pinning.
As shown in Fig. \ref{Fig1}, the value of $v_T$ deduced from Eq. \ref{eq:depinning_T} is found to coincide with the linear extrapolation of the flow regime observed at high drive. In order to test the generality of this observation, we have analyzed the flow regime for different material and temperature. The Table~\ref{table:table1} reports the value of DW mobility $m_{fl}$ deduced from a linear fit of the linear flow regime, which was only observed for Pt/Co/Pt and (Ga,Mn)As films. In Fig. \ref{fig:R_v(Hd)_Hd}, we report the variation of the ratio $v_T[(H_d,T)/m_{fl}(T)$ with the depinning field $H_d$. As it can be observed, all the points collapse on a single line. The slope is equal to 1, which indicates that the depinning velocity can be written $v_T(H_d,T)=m_{fl}(T)H_d$. We can deduce that the depinning velocity $v_T$ correspond to the flow velocity that DW would have in the absence of pinning.  Following Eq. \ref{eq:depinning_T}, this suggests that the DW velocity of the glassy dynamics to scale with the domain wall width parameter $\Delta$ and damping parameter $\alpha$ following Eq. \ref{eq:mobility}.
\\




\section{Conclusion}
\label{sec:conclusions}

In conclusion, we propose a quantitative and comparative study of domain wall pinning in different ferromagnets. The latter is based on a clear discrimination between universal and material dependent behaviors of the creep and depinning dynamical regimes. The determination of effective pinning pinning parameters allows to explore the interplay between micromagnetic and pinning properties of ferromagnets and domain wall dynamics.

Our works opens a way for a better understanding of the microscopic origin of pinning in magnetic systems. 
Our analysis provides a functional dependency of DW glassy velocity with the micromagnetic and pinning parameters. 
This should have important implications for a comparison between experimental and theoretical studies as micromagnetic simulations on one hand~\cite{voto_jpdap_2016} and to statistical models for interface motion in disordered elastic systems~\cite{agoritsas_PRE_13,Ferrero_CRP_2013} in the other.
In particular, the latter allows a more stringent test for the different equilibrium and depinning universality classes proposed to describe the non-equilibrium dynamics at different length-scales and velocities.
Moreover our analysis should help for a better understanding of chiral effects on DW dynamics~\cite{je_prb_2013_DMI,lavrijsen_prb_2015_PtCoPt_in-plane,pellegren_prl_2017} and in particular the contribution of the Dzyaloshinskii-Moriya interaction. They 
should manifest through the fundamental field, temperature and velocity scales controlling the macroscopic universal response of DWs.

\begin{acknowledgments}
We wish to thank E. Agoritsas and J. Curiale for fruitful discussions, J.-P. Adam, L. Herrera Diez, D. Ravelosona, and N. Vernier for sharing their data on CoFeB. S. B. and V. J. acknowledge support by the French-Argentina project ECOS-Sud num. A12E03. This work was also partly supported by the Labex NanoSaclay, Ref.: ANR-10-LABX-0035. S. B. and A. B. K. acknowledge partial support from project PIP11220120100250CO (CONICET).
\end{acknowledgments}

%
%
%
%
%
%
\section*{\bf Annex I: Micromagnetic and pinning parameters}

Here, we discuss technical details on the determination of micromagnetic parameters, which are listed in Table \ref{table:table1} (see Annex II) together with the fundamental pinning parameters controlling domain wall dynamics.
The values of magnetization saturation $M_s$, domain wall energy $\sigma$, and thickness parameter $\Delta$ are directly reported from the publications when they are available  or deduced from the relations $\sigma = 4\sqrt{AK_{eff}}$ and $\Delta =\sqrt{A/K_{eff}}$, where $A$ is the stiffness energy and $K_{eff}$ the effective anisotropy.

For Pt/Co/Pt, the data were taken from Refs. \onlinecite{metaxas_PRL_07_depinning_thermal_rounding} (different thicknesses $t$, and room temperature), \onlinecite{gorchon_PRL_2014} (thickness $t= 0.45$~nm and variable temperature) and \onlinecite{diaz_PRB_2017_depinning} (thickness $t= 0.5$~nm and variable temperature). For $0.5$~nm thick Pt/Co/Pt film\cite{diaz_PRB_2017_depinning}, the room temperature micromagnetic parameters $M_s$, $\sigma$ and $\Delta$ correspond to those of the $0.5$~nm thick film of Ref. \onlinecite{metaxas_PRL_07_depinning_thermal_rounding}.
 As proposed in Ref.~\onlinecite{gorchon_PRL_2014}, the thermal dependence of $M_s$ was deduced from polar magneto-optical Kerr rotation.  Since $A \sim M_s (T)^2$ and $K_{eff} \sim M_s (T)$, we assumed the following temperature variations for the DW energy $\sigma \sim M_s (T)^{3/2}$ and the domain wall width parameter $\Delta \sim M_s (T)^{1/2}$.

For Au/Co/Au, $M_s$ was taken equal to its bulk value\cite{wohlfarth1980ferromagnetic}.

For (Ga,Mn)(As,P), the temperature variation of the saturation magnetization $M_s$ was deduced from PMOKE measurement and was found similar to that observed for (Ga,Mn)As\cite{thevenard_prb_2011,dourlat_prb_2008}. The Curie temperature is $74 \pm 1$~K. According to the concentration of Mn atoms, we assumed $M_s(T=0$~K$)=40$~kA/m \cite{haghgoo_prb_2010}. The domain wall width parameter $\Delta$ was deduced from the slope ($m_{fl}$) of the precessional flow regime and the prediction of the one dimensional model: $m_{fl}=v/H=\alpha \gamma \Delta /(1+ \alpha^2)$, with $\gamma=1.76 10^{11}$~Hz/T and $\alpha$=0.3\cite{dourlat_prb_2008,thevenard_prb_2011}. The obtained value for $\Delta=11.5 \pm 0.5$~nm is almost temperature independent and close to the value reported in Ref. \onlinecite{haghgoo_prb_2010} ($\Delta=8 \pm 1$~nm).

For CoFeB/MgO with different Co and Fe concentrations, the data were taken from Ref. \onlinecite{Burrows_APL_13_CoFeB}. The DW energy is deduced from $\sigma = 4K_{eff}\Delta$, with $K_{eff}=M_s H_{k,eff}/2$.

For CoFeB/MgO with different irradiation dose, the data were taken from Ref. \onlinecite{herrera-diez_apl_2015}.

For the [Co/Ni] superlattices, data were taken from Refs. \onlinecite{legall_apl_2015} and \onlinecite{yamada_apex_2011}. For the DW energy, we used $\sigma = 4K_{eff}/\Delta$. For the stiffness energy, we took the value ($A=$10~pJ/m) reported in Ref. \onlinecite{yamada_apex_2011}.

%

\section*{\bf Annex II: Tables}
\begin{table*}[h!]
\centering
\begin{ruledtabular}
\begin{tabular}{|l|c|r|}

  \bf{Characteristic length-scale} & \bf{Variable} & \bf{Name}\\  	 
  \hline
	 \it{Micromagnetic continuum description} 
      & $A$         & stiffness energy\\
	  & $K_{eff}$   &anisotropy energy\\
	  & $M_s$       &saturation magnetization\\
	  & $\alpha$    &Gilbert damping factor\\
      & $\gamma$    &gyromagnetic factor\\      
  \hline
	\it{Microscopic domain wall scale $L<L_c$} 
      & $\Delta \sim \sqrt{\frac{K_{eff}}{A}}$         &domain wall width parameter\\
	  DW micromagnetic structure
      & $\sigma \sim \sqrt{K_{eff}A}$   &domain wall energy\\
	  & $m_{fl}=\frac{\gamma \Delta \alpha}{1+\alpha^2}$       &domain wall mobility\\
      
	  DW Pinning
      & $D$   &strength of the pinning disorder\\
      & $\xi$  &correlation length of the disorder\\
      & $f_{pin} \sim \frac{\sqrt{D}}{\xi}$  &pinning force (cf. Eqs. \ref{eq:larkin0_agoritsas} and~\ref{eq:ksi0_agoritsas})\\
\hline
	  \it{Mesoscopic DW scale $L \sim L_c$} 
       &$T_d$& depinning temperature (cf. Eqs. \ref{eq:v-creep} and \ref{eq:E-creep})\\
       \it{(Larkin Regime)}
      &$H_d$& depinning threshold (cf. Eqs. \ref{eq:v-creep} and \ref{eq:E-creep})\\
      &$v_T$& depinning velocity (cf. Eq. \ref{eq:depinning_T})\\
\hline
	  \it{Macroscopic DW scale $L \gg L_c$}
      & $\mu$         & creep exponent (cf. Eq. \ref{eq:E-creep})\\
(\it{Random Manifold regime}) Creep regime
      & $\Delta E/(k_BT_d)$   &universal energy barrier function (cf. Eq. \ref{eq:E-creep})\\
      
	  Depinning transition
      &$v(H_d)$& velocity at depinning (cf. Eqs. \ref{eq:v-creep} and \ref{eq:E-creep})\\      
      & $\beta$   &depinning exponent (field effects) (cf. Eq. \ref{eq:depinning_H})\\
	  & $\psi$       &depinning exponent (thermal effects) (cf. Eq. \ref{eq:depinning_T})\\
	  & $x_0$    &universal metric factor (cf. Eq. \ref{eq:depinning_H})\\
      & $g(x/x_0)$     &universal function of depinning (cf. Eq. \ref{eq:g_function})\\
\end{tabular}
\caption{\label{table:table2}  \textbf{Parameters describing domain wall dynamics}, classified according to the length-scale they emerge.}
\end{ruledtabular}
\end{table*}

\begin{table*}[h!]
\centering
\begin{ruledtabular}
\begin{tabular}{|l|c|c|c|r|r|r|r|r|r|}

  
  
   Material & \bf{$t (nm)$} & \bf{$T(K)$} & \bf{$   T_d(K)$} & \bf{$   H_d(mT)$} & \bf{$   v(H_d)(m/s)$} & \bf{$ m_{fl}(m/(s.mT))$} &\bf{$   M_s(kA/m)$} & \bf{$   \sigma (\mu J/m^2)$} & \bf{$\Delta(nm)$}\\
  	 
  \hline
	Pt/Co/Pt 
	  & 0.5 & 293 & 2558(10) & 28.5(2) & 5.7(0.2) & 0.276(0.005)& 910& 9030& 6.2\\
	 Ref.~\onlinecite{metaxas_PRL_07_depinning_thermal_rounding} 
	  & 0.6 & & 4145(25) & 56(1) & 10.6(1.0)& 0.325(0.005)& 1130& 11700& 5.5\\
	  & 0.7 & & 6490(30) & 76(1) & 16.6(1.0)& 0.370(0.005)& 1200& 10700& 6.7\\
	  & 0.8 & & 9720(45) & 72(1) & 18.4(1.0)& 0.454(0.005)& 1310& 10200& 8.6\\
	irradiated & 0.5 & & 2260(50) & 15(1) & 7.5(1.0)& 0.676(0.005)& 700 & 3080& 15.6\\
	
	\hline
	Pt/Co/Pt 
	 & 0.45 & 293 & 1900(100) & 91(4) & 59(2) & & 800& 7400 & 5.8\\
	 Ref.~\onlinecite{gorchon_PRL_2014} 
	  & & 200& 2500(100) & 110(5) &61(2)& & 1120& 12300& 6.9\\
	  & & 150 & 2700(100) & 135(5) & 90(2)&& 1260& 14700& 7.3\\
	  &  &100 & 3200(100) & 141(5) & 75(2)&& 1370& 16700& 7.6\\
	  &  &50 & 3500(100) & 150(5) & 81(2)& & 1470& 18500& 7.9\\
		 	
	\hline
	Pt/Co/Pt 
      & 0.5 & 293 & 2650(50) & 57(3) & 12.0(0.5) & 0.288(0.005)& 910& 9030& 6.2\\
	Ref.~\onlinecite{diaz_PRB_2017_depinning}
	  &  & 225 & 2750(50) & 75(3) & 14.8(0.5) & 0.307(0.005)& 1120& 12300& 6.9\\
	  &  & 150 & 2700(50) & 101(3) & 18.6(0.5) & 0.280(0.005)& 1330& 16000& 7.5\\
	  &  & 100 & 3090(50) & 107(3) & 18.7(0.5) & 0.260(0.005)& 1470& 18500& 7.9\\
	  &  & 50  & 2860(50) & 120(3) & 19.5(0.5) & 0.292(0.005)& 1600& 21000& 8.2\\
	  &  & 10  & 660(50) & 130(3) & 21.0(0.5) & 0.363(0.005)& 1720& 23500& 8.5\\
	  &  & 4.4 & 450(50) & 136(3) & 20.0(0.5) & 0.364(0.005)& 1730& 23700& 8.5\\

	\hline
	Au/Co/Au &  1.0 & 318 & 28400(1500) & 88.0(1.0) & 8.7(1.0)& &1400 & &\\
	Ref.~\onlinecite{kirilyuk_JMMM_97_AuCoAu} & & 273 & 29000(1500) & 96.5(1.0) & 8.4(1.0)& &1400 & &\\
	 & & 243 & 29300(1500) & 102.5(1.0) & 7.3(1.0)& &1400 & &\\
	 & & 213 & 29400(1500) & 110.0(1.0) & 7.5(1.0)& &1400 & &\\ 
	 & & 183 & 28800(1500) & 115.0(1.0) & 7.4(1.0)& &1400 & &\\
	 & & 150 & 25800(1000) & 122.6(1.0) & 9.7(1.0)& &1400 & &\\
	 
  \hline
  Co$_{20}$Fe$_{60}$B$_{20}$ an & 1 & 293 & 1800(100) & 6.6(0.2) & 1.7(0.5)& & 1100& 9200& 9.7\\
  Co$_{20}$Fe$_{60}$B$_{20}$ ag       & & & 1800(100) & 4.8(0.2) & 2.3(0.5)& & 1000& 2700& 30.2\\
  Co$_{40}$Fe$_{40}$B$_{20}$ an       & & & 1400(100) & 5.0(0.2) & 2.0(0.5)& & 880& 7400& 10.7\\
  Co$_{40}$Fe$_{40}$B$_{20}$ ag       & & & 2000(100) & 4.3(0.2) & 4.6(0.5)& & 1100& 4900& 21\\
  Co$_{60}$Fe$_{20}$B$_{20}$ an       & & & 2200(100) & 3.5(0.5) & 2.3(0.5)& & 1100& 5100& 27.7\\
  Ref.~\onlinecite{Burrows_APL_13_CoFeB} & && && && &&\\

   \hline
  Co$_{20}$Fe$_{60}$B$_{20}$ an &1 &293 & & & & & & &\\
  Ref.~\onlinecite{herrera-diez_apl_2015} & && && && &&\\ 
  Dose $\times 10^{19}$He/nm$^2$ & & & & & & & & &\\
     0& & & 2640(100) & 10.5(0.2) & 2.1(0.5)& & 880& &\\
   0.1& & & 2580(100) & 10.2(0.2) & 2.4(0.5)& & 860& &\\
   0.2& & & 2570(100) & 10.9(0.2) & 2.4(0.5)& & 890& &\\
   0.4& & & 2640(100) & 11.4(0.2) & 3.0(0.5)& & 760& &\\
   0.6& & & 2500(100) & 11.4(0.5) & 3.2(0.5)& & 810& &\\
   0.8& & & 2510(100) & 17(0.2) & 3.8(0.5)& & 840& &\\
   1  & & & 2540(100) & 22.4(0.2) & 3.3(0.5)& & 770& &\\
   1.2& & & 2670(100) & 23.0(0.2) & 2.9(0.5)& & 710& &\\
   1.4& & & 2680(100) & 25.0(0.5) & 2.8(0.5)& & 680& &\\
   1.6& & & 2300(100) & 46.3(0.5) & 1.7(0.5)& & 670& &\\

  \hline
	TbFe 
	& 5$\times$1.8 & 271 & 5750(50) & 295(5) & 1.4(0.1)& & & &\\
	Ref.~\onlinecite{jeudy_PRL_2016_energy_barrier} & & 289 & 4200(50) & 225(5) & 1.8(0.1)& & & &\\
	 & & 304 & 3050(50) & 130(5) & 1.7(0.1)& & & &\\
	 & & 310 & 2600(50) & 100(5) & 1.7(0.1)& & & &\\
	 & & 315 & 2200(50) & 80(5) & 1.8(0.1)& & & &\\
	 
   \hline
  (Ga,Mn)(As,P)
     & 12 & 10 & 616(10) & 6.2(0.1) & 1.8(0.1) & 0.537(0.005)& 38& 130& 11.1\\
	 Ref. \onlinecite{jeudy_PRL_2016_energy_barrier}
	 & & 30 & 1440(20) & 5.8(0.1) & 1.8(0.1)& 0.564(0.005)&  34& 100& 11.6\\
	 & & 50 & 1140(20) & 5.6(0.1) & 2.0(0.2)& 0.566(0.005)&  26& 60& 11.7\\
	 & & 65 & 815(10) & 5.5(0.1) & 2.3(0.1)& 0.58(0.01)& 18& 30& 12.0\\
  
  \hline
	[Co/Ni] superlattice &  1.1$\times$3 & 293 & 51300(500) & 25(1) & 5.1(0.1)& &930 & 6900&5.8\\
	Ref. \onlinecite{legall_apl_2015}& && && && &&\\
	
   \hline
   [Co/Ni] superlattice &  1.2$\times$4 & 293 & 30000(10000) & 12.5(2.0) & 7.5(2.5)& &680 &8300 &6.95\\
	Ref. \onlinecite{yamada_apex_2011}& && && && &&\\
\end{tabular}
\caption{\label{table:table1}  \textbf{Material and temperature dependent parameters.}
For each material, the thickness ($t$) and the temperature of the experiment ($T$ (K)) is indicated. The depinning temperature ($T_d$) and magnetic field ($H_d$) and the domain wall velocity at the depinning field ($v(H_d)$) are deduced from a fit of the velocity curves (see text). $m_{fl}$~(m/(s.mT)) is the best fit for the slope of the linear precessional regime. The saturation magnetization $M_s$, the DW energy $\sigma$ and the DW thickness parameter $\Delta$ are  extracted or deduced from the references indicated below the name of material.} 
\end{ruledtabular}
\end{table*}

\bibliography{refs_DW_pinning}
\end{document}